\documentclass[unsortedaddress,
twocolumn,showpacs,preprintnumbers,amsmath,amssymb]{revtex4}

\usepackage{graphicx}
\usepackage{dcolumn}
\usepackage{bm}
\usepackage{latexsym}

\begin{document}

\title{Platoon formation in a traffic model with variable anticipation}

\author{M.E. L\'{a}rraga}
\email{mel@cie.unam.mx}
\affiliation{Centro de Investigaci\'on en Energ\'{\i}a, UNAM, 
A.P. 34, 62580 Temixco, Mor. M\'exico}

\author{J.A. del R\'{\i}o}
\email{antonio@servidor.unam.mx}
\affiliation{Centro de Investigaci\'on en Energ\'{\i}a, UNAM,
A.P. 34, 62580 Temixco, Mor. M\'exico}

\author{A. Schadschneider}
\email{as@thp.uni-koeln.de}
\affiliation{Institute for Theoretical Physics, Universit\"at zu K\"oln, 
50937 K\"oln, Germany}

\date{\today}%

\begin{abstract}
A cellular automaton model of traffic flow taking into
account velocity anticipation is introduced.  The strength of
anticipation can be varied which allows to describe different
driving schemes. We find phase separation into a free-flow regime
and a so-called $v$-platoon in an intermediate density regime. 
In a $v$-platoon all cars move with velocity $v$ and have vanishing 
headway. The velocity $v$ of a platoon only depends on the strength of
anticipation. At high densities, a congested state characterized
by the coexistence of a $0$-platoon with several $v$-platoons is reached.
The results are not only relevant for automated highway 
systems, but also help to elucidate the effects of anticipation
that play an essential role in realistic traffic models.
From a physics point of view the model is interesting because
it exhibits phase separation with a condensed phase in which
particles move coherently with finite velocity coexisting with
either a non-condensed (free-flow) phase or another condensed
phase that is non-moving.
\end{abstract}

\maketitle


\section{Introduction}

In the last years, the continuous increase of traffic demand  has 
prompted authorities around the world to place more emphasis on
 improving the efficiency and capacity on the roadway systems. 
Ecological considerations, space and budgetary constraints have 
limited solving traffic congestion by upgrading and constructing 
new roadway systems. Advanced technologies for vehicular traffic 
have been developed as a mean to improve the management of existing 
system and thus to solve traffic congestion, environmental issues 
and improve traffic safety. However, to achieve these aims, 
an accurate forecast of the impact of these technologies is
critical before their final deployment.

Testing these advanced technologies on real traffic is not always 
feasible. In contrast, computer simulations as means for evaluating 
control and management strategies in traffic systems have gained 
considerable importance because of the possibility of taking into 
account the dynamical aspects of
traffic (see for example, \cite{EKV95,GETRAM,autobahn}) and assessing the 
performance of a given advanced technology in a short time.

Cellular Automata (CA) models for traffic flow \cite{a5,ito} have shown
the ability to capture the basic phenomena in traffic flow \cite{San}.
Cellular automata are dynamic models in which space, time and state
variables are discrete.  Discrete space consists of a regular grid of
cells, each one of which can be in one of a finite number $k$ of
possible states. All cells are updated in discrete time-steps.  The
new state of a cell is determined by the actual state of the cell
itself and its neighbor cells. This local interaction allows us to
capture micro-level dynamics and propagates them to macro-level
behavior. The discrete nature of CA makes it possible to simulate
large realistic traffic networks using a microscopic model faster than
real time \cite{ito,nagelreview,es}.  Now, almost eleven years after the
introduction of the first CA models, several theoretical studies and
practical applications have improved the understanding of empirical
traffic phenomena (see e.g.\ \cite{a2,San,and,ker}). Moreover, CA models
have proved to be a realistic description of vehicular traffic in
dense networks \cite{nagelreview,es}.

In this paper, we propose a single-lane probabilistic model based
on the first CA model of Nagel and Schreckenberg \cite{a5} (hereafter
cited as the NaSch model) to describe the effects of several
anticipation schemes in traffic flow. Anticipation in traffic means
that drivers estimate their leader's velocities for future time-steps
\cite{knospe2lane,knospe,chen,LWK,schneider,knospe2,wagner}.
This can lead to an increase of the vehicular capacity and a decrease
of the speed variance. However, incorporating different driving
strategies requires a safety distance with respect to the preceding
vehicles. For that purpose we introduce a new parameter in the
deceleration process, called \emph{anticipatory driving parameter}, 
to estimate the velocity of the precedent vehicle.  This
estimation, plus the real spatial distance to the leading vehicle,
determine a safe braking distance.  By
appropriately tuning this new parameter different traffic situations
of non-automated, automated and mixed traffic can be considered.
Furthermore, the anticipatory driving parameter is relevant for
Automated Highway Systems (AHS) \cite{AH-JVSD99a,Larraga}.

According to simulation results from our proposed model, the relations
derived from the density vs. velocity and density vs.\ flow curves are
in agreement with the fundamental diagrams that describe these
relations in real non-automated traffic. In addition,
simulation results from our model in the case of high anticipation
(like automation) describe one of the interesting phenomena in traffic
flow, \emph{formation of platoons}. We will show that, in contrast to
models without anticipation, {\em dense} platoons can be formed where
all cars move coherently with some finite velocity $v>0$.  The mechanism for
platoon formation is not only of great importance for AHS to increase
highway capacity in a much safer way \cite{SCHI94}, but also helps to
understand the essential role of anticipation effects in realistic
traffic models. By varying the anticipatory driving parameter three
different regimes, characterized by different slopes of the
fundamental diagrams, can be observed.  Apart from a free-flow and a
congested phase, an additional regime where platoons of cars moving
with the same velocity $v<v_{max}-1$ exist.

The paper is organized as follows. In Section~\ref{modeldef} we define a
modified NaSch model to consider different driving strategies. In this
way, we introduce a new parameter to determine a velocity-dependent
braking distance. In Section~\ref{simuresults}, we present the results 
of our investigations. We show results for the fundamental diagram and
different values of the anticipation parameter. A comparison with
those for other models and real non-automated traffic is presented.
Phase separation into a free-flow regime and so-called $v$-platoon is
observed in a certain intermediate density regime. 
For large densities, in the congested state phase separation into 
a dense jam ($0$-platoon) and $v$-platoons is observed.
The flow structure determined by the
existence of dense platoons with velocity $v$ is calculated.
Dependence of an optimal anticipation level on the density is found.
Analytical results are in excellent agreement with results from
computer simulations. In the concluding Section~\ref{summary} we 
summarize our results and discuss the relevance of our results for
traffic models and real traffic.


\section{Definition of the model}
\label{modeldef}

In this section we introduce the proposed model. It is defined on a
one-dimensional lattice of $L$ cells with periodic boundary conditions,
which corresponds to a ring topology with the number of vehicles preserved.
Each cell is either empty, or it is occupied by just one vehicle traveling
with a discrete velocity $v$ at a given instant of time. All vehicles have
a velocity that ranges from $0,\ldots, v_{max}$. In addition, and for
simplicity, only one type of vehicles is considered. The time-step ($\Delta t
$) is taken to be one second, therefore transitions are from $t\rightarrow
t+1$. It can also be easily modified.

Let $v_{i}$ and $x_{i}$ denote the current velocity and position,
respectively, of the vehicle $i$, and $v_{p}$ and $x_{p}$ be the
velocity and position, respectively, of the vehicle ahead (preceding
vehicle) at a fixed time; $d_{i}:=x_{p}-x_{i}-1$ denotes the distance
(number of empty cells) in front of the vehicle in position $x_{i}$ that
sometimes is called headway.

The dynamics of the model are defined by the following set of rules, 
that are applied to all $N$ vehicles on the lattice each time-step:
\begin{description}
\item[R1:] Acceleration \\ 
If $v_{i}<v_{max}$, the velocity of the car $i$ is increased by one, i.e.,
$$
v_{i}\rightarrow \min(v_{i}+1,v_{max}).
$$
\item[R2:] Randomization \\
If $v_{i}>0$, the velocity of car $i$
\ is decreased randomly by one unit with probability $R$, i.e.,
$$
v_{i}\rightarrow \max(v_{i}-1,0) \quad {\rm with\ probability}~R.
$$
\item[R3:] Deceleration \\
If $d_{i}^{s}<v_{i}$, where 
$$
d_{i}^{s}=d_{i}+\left[(1-\alpha )\cdot v_{p}+\frac{1}{2}\right],
$$ 
with a parameter $0\leq \alpha \leq 1$, the velocity of 
car $i$\ is reduced to $d_{i}^{s}$. $[x]$ denotes the integer
part of $x$, i.e.\ $[x+\frac{1}{2}]$ corresponds to rounding $x$ to
the next integer value.\\
The new velocity of the vehicle $i$ is therefore
$$
v_{i}\rightarrow \min(v_{i},d_{i}^{s}).
$$
\item[R4:] Vehicle movement \\
Each car is moving forward according
to its new velocity determined in steps 1-3, i.e.,
$$
x_{i}\rightarrow x_{i}+v_{i}.
$$
\end{description}
Rules $R1$, $R2$ and $R3$ are designed to update velocity of vehicles;
rule $R4$ updates position. According to this, state updating is
divided into two stages, first velocity, second position. Note that
this division follows the scheme in differential equation integration
that first updates the time derivative and then the value of the
state. It is important to mention that we are changing the order of
the rules in comparison with NaSch model since $R2$ is
applied before $R3$.

Rule $R1$ indicates that all the drivers would like to reach the maximum
velocity when possible. Rule $R2$ takes into
account the different behavioral patterns of the individual drivers  in
which with no apparent reason a driver decreases its speed. These situations
include, for example, cases of overreaction in braking or incidents along
the highway that distract drivers, and random fluctuations.

Rule $R3$ is the main modification to the original NaSch model \cite{a5}. In
this rule the distances between the  $ith$ and $(i+1)th$ vehicles, and their
corresponding velocities are considered. Knowledge of the preceding
vehicle's velocity is incorporated through the \emph{anticipatory driving 
parameter} $\alpha $ with range $0\leq \alpha \leq 1$. 
Notice that, by only varying the parameter $\alpha$ in the term 
$d_{i}^{s}=d_{i}+[(1-\alpha ) v_{p} +1/2]$, different anticipatory 
driving schemes that require different safe braking distance with respect 
to the preceding vehicle can be modelled. 
If $\alpha $ takes its maximum value ($\alpha =1$) the speed of the
vehicle ahead is not considered in the deceleration process, i.e.\
anticipation is not considered. On the contrary, when $\alpha =0$ the speed
of the vehicle ahead is considered without restrictions, i.e., without
establishing a braking distance with respect to the precedent vehicle
\cite{Larraga}. This last case occurs
with either a very aggressive driver or when vehicles can obtain
information about the velocity of vehicles ahead \footnote{This 
will happen in Automated Highway Systems or in vehicles equipped with
appropriate sensors \cite{SCA}.} to allow small distances between vehicles 
(e.g.\ of the order 1~m). Intermediate values for $\alpha $ thus
represent different braking spacing policies or degrees of automation in the
vehicles or anticipatory driving schemes. Platooning schemes 
\cite{SCHI94}
imply values of $\alpha $ closer to zero and demand additional requirements
to preserve safety, like coordinated braking \cite{AH-JVSD99a}.
Independent vehicle driving with low level of anticipation implies values 
of $\alpha $ closer to 1 in order to preserve safety levels: the larger 
$\alpha$ is the larger braking distance is. 
Note that in order to determine $v_{n}$ consistently for all vehicles in
the case of periodic boundary conditions, rule $R3$ must be iterated at 
most $(v_{max}-1)$ times.
In real situations, the drivers always estimate the velocity of 
preceding vehicle and according to this and their way to drive
(relaxed or aggressive behavior) they choose a safe headway distance
to drive. Variation of $\alpha$ allows also to model these aspects.

Thus, the proposed model is able to represent different anticipatory 
driving schemes, and model the minimum braking distance required with only
one parameter $\alpha$, here referred to as \textit{anticipatory driving 
parameter}.

We emphasize that the CA model as presented here is a minimal 
model in the sense that all four steps $R1$-$R4$ are
necessary to reproduce the basic features of real traffic, however,
additional rules may be needed to capture more complex situations 
\cite{knospe}.


\section{Simulation Results}
\label{simuresults}

To simulate the CA model proposed in the previous section, the typical
length of a cell is around $7.5$~m. It is interpreted as the length of
a vehicle plus the distance between cars in a dense jam, but it can be
suitably adjusted according to the problem under consideration. With
this value of the cell size and a time-step of 1 s, $v=1$ corresponds
to moving from one cell to the downstream neighbor cell in one
time-step, and translates to $27~$km/h in real units.  The maximum
velocity is set to $v_{max}=5$, equivalent to $135$~km/h.  The
total number of cells is assumed to be $L=10^{4}$, and the density $\rho$
is defined as $\rho =N/L$, where $N$ is the number of cars on the
highway.  Initially, $N$ vehicles are distributed randomly on the lane
around the loop with an initial speed taking a discrete random value
between 0 and $v_{max}$.  Since the system is closed, the density
remains constant with time.

Velocities are updated according to the velocity updating rules $R1-R2-R3$ 
and then all of cars are moved forward in step $R4$.  Each
run is simulated for $T=6L$ time-steps. In order to analyze results,
the first half of the simulation is discarded to let transients die
out and the system reach its steady state. For each simulation a
value for parameter $\alpha $ is established by taking into account
the desired anticipation degree and thus, controlling the safe
braking distance among vehicles. In the following, the value of
$\alpha$ is the same for all vehicles (homogeneous drivers).


\subsection{Comparison with real non-automated traffic}

\begin{figure}[htb]
\centering
\includegraphics[width=0.8\columnwidth,angle=270]{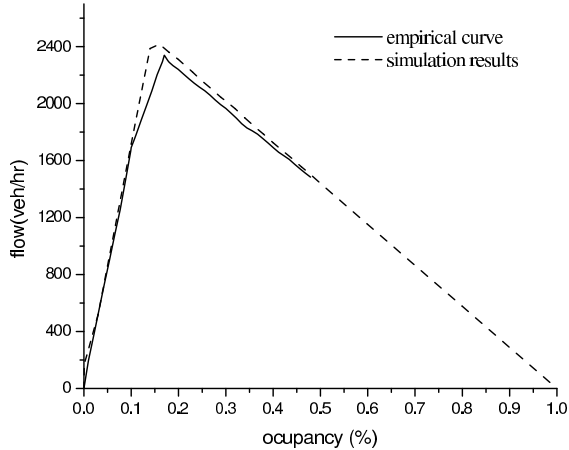}
\caption{Typical form of an empirical fundamental diagram
taken from \cite{Hall} (solid line) in comparison
with simulations of the proposed model for $R=0.2$ and
$\alpha=0.75$ (dashed line).}
\label{f1}
\end{figure}

The fundamental diagram is one of the most important criteria to show
that the model reproduces traffic flow behavior. This diagram
characterizes the dependence of the vehicles flow on density. We
obtain a fundamental diagram for the proposed model with $R=0.2$ and
$\alpha=0.75$, see Fig.~\ref{f1}.  This $\alpha$ value corresponds to
cautious estimation of the preceding car's velocity.

As we can see from Fig.~\ref{f1}, the curve of the proposed model
(dashed line) is consistent with the characteristic curve of the
measured fundamental diagram (solid line) taken from \cite{Hall}.  The
critical density and the maximum flux of our model are $(\rho
_{c},q_{max})=(16\%,$ $2417$\ cars/h), closer to the empirical curve
values of $(\rho _{c},$ $q_{max})=(17\%,$ $2340$\ cars/h), in
comparison with other existing models, see
\cite{LWK}.  Note the quantitative agreement between simulated and
empirical curve of the fundamental diagram in its decreasing part.

Moreover, as we can see from Fig.~\ref{f1}, a decrease in the
experimental curve slope is observed at $(10\%,$ $1692$~cars/h),
corresponding to a reduced mean velocity of vehicles near the critical
density. Thus, we make a small modification to the proposed model
based on an idea of \cite{Eme}.
We modified the deceleration rule such that a vehicle can
only reach the maximum velocity if its braking distance to the
precedent vehicle is greater than nine cells ($67.5$~m, i.e, the
density is lower than $0.1$).
The modified rule $R3$ of the model is then as follows:
\begin{description}
\item[R3':] If $v_{i}=5$ (i.e.\ $v_i=v_{max})$ and $d_{i}^{s}\leq 9$~cells 
then
$$v_{i}\rightarrow \min(v_{i}-1,d_{i}^{s})\phantom{aaaaaaaaaaaaaaaaaaaaaaaa}$$
else (like rule \emph{R3} of our original model)
$$v_{i}\rightarrow \min(v_{i},d_{i}^{s}).\phantom{aaaaaaaaaaaaaaaaaaaaaaaa}$$
\end{description}

\begin{figure}[tbh]
\centering
\includegraphics[width=0.8\columnwidth,angle=270]{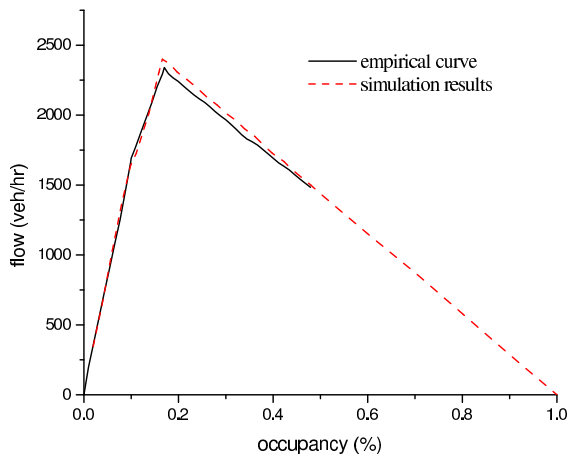}
\caption{Fundamental diagram for the \textit{modified model} (\emph{R3'}). 
This
diagram is in agreement with the experimental curve in a quantitative way
($R=0.2 $ and $\alpha = 0.75 $).}
\label{f2}
\end{figure}

In Fig.~\ref{f2}, the results obtained from our modified model are
shown and compared with the experimental curve from real traffic data
\cite{Hall}. The curve resulting from the modified model fits
the real curve in both its increasing part and its decreasing part
quite well. Thus, the results from the modified model agree quantitatively
with the experimental shape of the fundamental diagram. This is an
indication that in real traffic the value of $\alpha$ may depend
on velocity.

In this subsection, we have shown that the given model can reproduce some
common characteristics of the real manual (non-automated) traffic
flow.  However, with the new parameter it is possible to consider
several anticipation schemes for traffic flow: non-automated, mixed
and automated traffic flow. This will be shown in the following
for the original model (\emph{R3}) defined in Sec.~\ref{modeldef}.


\subsection{Modeling different anticipation schemes}

\begin{figure}[htb]
\centering
\includegraphics[width=0.8\columnwidth,angle=270]{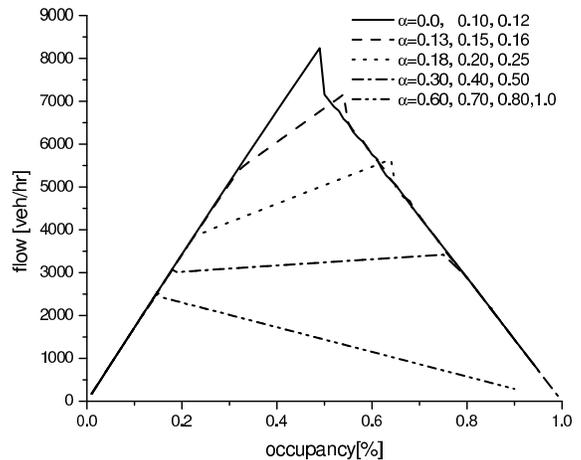}
\caption{Fundamental diagram for different values of the anticipation
parameter $\alpha$ and $R=0.2 $.}
\label{f3b}
\end{figure}

Determination of the impact of different driving strategies is important 
in order to propose automated traffic alternatives. 
Following that proposal, we decided to investigate traffic flow 
behavior using our model. As mentioned above, the parameter $\alpha $ 
represents 
the way in which different driving strategies adopt a braking distance 
with respect to the preceding vehicles.
Varying the parameter $\alpha $, these strategies can be tuned. 
In Fig.~\ref{f3b} we show the fundamental diagram of the proposed model
 with a fixed value of $R=0.2$ and different values of $\alpha $. 
From this diagram, the impact of the driving strategies coded in $\alpha $
can be observed. Smaller values of 
$\alpha$ imply larger flows, that is, higher levels of anticipation. 
Here vehicles keep a less safe braking distance, leading to an increase in 
the vehicular capacity. This behavior is in agreement with, for example, 
platooning strategies that exploit the knowledge of the velocity of 
precedent vehicles and require a smaller distance among vehicles (near 1~m), 
so increasing the flow.

\begin{figure}[htb]
\centering
\includegraphics[width=0.8\columnwidth,angle=270]{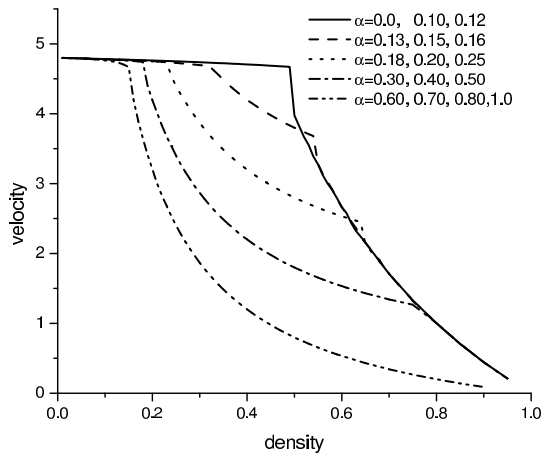}
\caption{Relationship between mean velocity and density for $R=0.2$ and
different values of $\alpha$.}
\label{f4}
\end{figure}

It is important to notice that for values of $\alpha$ from 0.13 to
0.50 a second positive slope corresponding to a mixed branch is
observed in the fundamental diagram. It is interesting
that the initial positive slope, corresponding to a free-flow region
where there are no slow vehicles, is similar for all values of
$\alpha$.  Here the vehicles travel at near maximum speed. For the
second branch, on the other hand, the flow is increased with
non-maximum velocity, indicating a mixed region due to anticipation
effects (Fig.~\ref{f4}). In order to analyze the role of the anticipation, 
we show the average velocity as function of the density  
for the same parameter values as in Fig.~\ref{f3b}.

As we can see from Fig.~\ref{f4}, higher levels of anticipation
(smaller values of $\alpha$) imply a larger density interval for the
free-flow region. For values $0.13 <\alpha< 0.50$, after the
free-flow region, traffic flow organizes in a so-called mixed region
with a lower mean velocity.  In this mixed region, in addition to free
flowing vehicles, and vehicles moving in platoons where all cars have
the same velocity and vanishing headway exist. The existence of this mixed
region indicates that a suitable estimation of the velocity
of precedent vehicle, coded in $\alpha$, allows that more cars to fit
on the road and the flow increases ever for values of large-density.

\begin{figure}[htb]
\centering
\includegraphics[width=0.8\columnwidth]{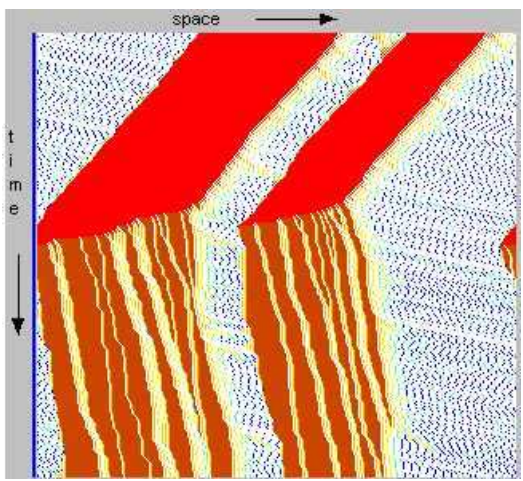}
\caption{Space-time diagram showing the time evolution of a system
simulated initially with $\alpha=0.51$. After switching to $\alpha=0.5$,
the behavior changes dramatically.}
\label{f5}
\end{figure}

Fig.~\ref{f5} shows a space-time diagram for $R=0.2$, $\rho = 0.5$ 
that exemplifies the dramatic changes in the microscopic structure
when changing the value of $\alpha$  \footnote{For a Java applet of the 
simulations, see {\tt http://www.cie.unam.mx/xml/tc/ft/arp/simulation.html}}. 
The simulation is started with $\alpha=0.51$ where the jamming regions 
travel backwards. However, after some time we switch to $\alpha=0.5$ and 
immediately observe a 
dramatic change in the slope of the congested regions. They are now
traveling forward. This behavior has been observed before in anticipatory 
modeling \cite{wagner}.
Such structured flow observed in space-time diagrams increases the 
highway capacity due to the space reduction among vehicles. In these
simulations we have found that the branches of congested or jammed flow
collapse to a single region as in the VDR model \cite{s2s}. 
These results are analogous to those for slow-to-start models, 
because effectively the outflow from a jam is reduced compared to the
maximal flow.

\begin{figure}[htb]
\centering
\includegraphics[width=0.8\columnwidth,angle=270]{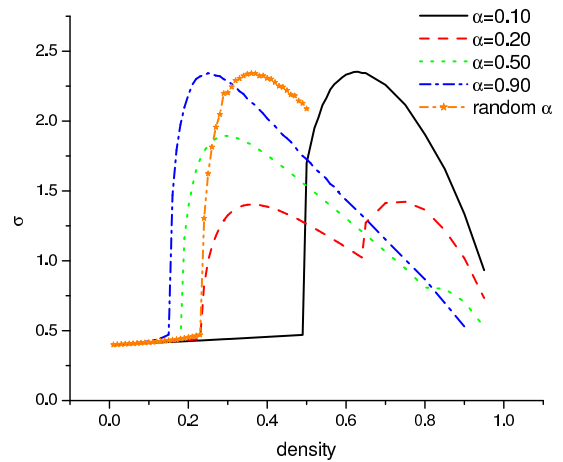}
\caption{Standard deviation of the speed.}
\label{f6}
\end{figure}

On the other hand, it is also important to analyze the efficiency of
traffic. In non-equilibrium situations, this is usually evaluated by
analyzing the entropy of the system. However, for non-physical systems
the analysis of the standard deviation of the main variable can give
equivalent information \cite{montroll}. In the context of traffic
flow, high standard deviation of speed means that, on average, a
vehicle would experience frequent speed changes per trip through the
system. In turn, the high speed variance could also increase the
probability of traffic accidents. Therefore, the standard deviation of
speed can be seen as an indicator of the efficiency in traffic flow.

In order to analyze the efficiency of traffic with different automation
levels, we calculate the standard deviation of speed (Fig.~\ref{f6}).  
For each value of $\alpha$ a maximum that occurs
shortly after the free-flow region can be clearly seen. In the
free-flow region, the speed variance is negligible since there are no
slow vehicles and fluctuations are extremely rare. Since the
free-flow region increases as $\alpha$ decreases, it is seems
reasonable to attempt traffic with the higher level of anticipation in
the range of density from 0 to 0.5. This selection not only produces a
state with higher flow, but also the lowest standard deviation, so
the efficiency is the highest. It can be clearly seen from Fig.~\ref{f6} 
that beyond the efficient density range, the level of
anticipation coded in $\alpha$ should be switched based on the density
regime: for $\rho \in(0.5,0.54]$, an efficient performance is found
with $\alpha=0.13$; however, for $\rho \in(0.54, 0.63]$ the highest
efficiency is attained with $\alpha=0.20$. Summarizing, the behavior
observed in Fig.~\ref{f6} indicates that 
the automation level should be determined depending on 
the density: higher densities require a lower automation level, that
is, a safer braking distance. 
  
Besides, we can also see from Figure \ref{f6}, the standard deviation
of speed resulting from assigning a random value of $\alpha$
between 0 and 1 to each vehicle. We stress that for each car, this value of
$\alpha$ is not changed during the time evolution.  Therefore, we simulate,
by means of random values of $\alpha$, traffic with
non-homogeneous drivers: aggressive, non-aggressive and relaxed
drivers. We can see from Fig.~\ref{f6} the region where the standard
deviation is negligible is close to that resulting from homogeneous
drivers with $\alpha=0.2$ (high-level anticipation). However, the
variance of vehicles speeds is $60\%$ larger than that corresponding to
homogeneous drivers.

\begin{figure}[htb]
\centering
\includegraphics[width=0.8\columnwidth,angle=270]{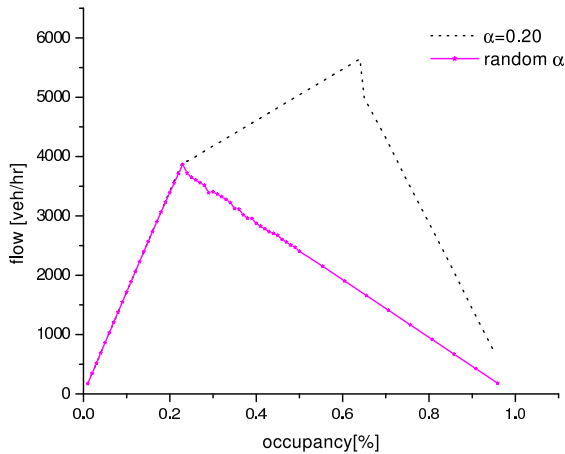}
\caption{Comparison of the fundamental diagrams of homogeneous systems 
with the random $\alpha$ case.}
\label{f7}
\end{figure}

In order to elucidate the effects of anticipation with non-homogeneous
drivers, the fundamental diagram obtained for this random $\alpha$
case is compared with that for $\alpha=0.2$ in a homogeneous
system (Fig.~\ref{f7}). Note that the second slope corresponding to
a mixed region is missing in the inhomogeneous system. 
In this case, the variance in the level of
anticipation considered by the drivers produces higher fluctuations of
speeds, and the flow decreases rapidly. Therefore some anticipation driving 
schemes have a strong impact on the efficiency of the system.  
All of these findings require physical explanations \cite{inprep}.

\subsection{Structure of the mixed states}
\label{sub-mixed}

The behavior in the mixed states is determined by the existence of dense 
platoons in which vehicles move coherently with the same velocity $v$.
In the following these will be denoted as $v$-platoons. The stationary
state then shows phase separation into a free-flow region and a
$v$-platoon. This is similar to the behavior observed in models
with slow-to-start rules where the system separates into free flow
and a dense jam, i.e.\ a $0$-platoon \cite{s2s}.

\begin{figure}[htb]
\centering
\includegraphics[width=0.8\columnwidth,angle=270]{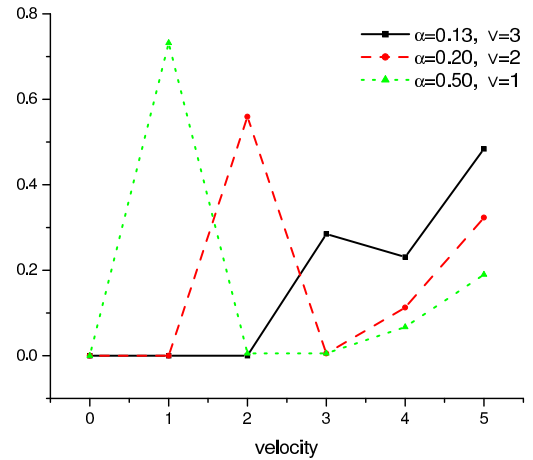}
\caption{Velocity distributions for density $\rho=0.4$ and different
values of $\alpha$.}
\label{fig_vel}
\end{figure}

Fig.~\ref{fig_vel} shows the velocity distributions of the different
branches. It can be clearly seen that only cars with velocity $v$
and free flowing cars (with velocity $v_{max}$ or $v_{max}-1$ due
to the randomization) exist.

Since the headway $d_i$ of a car $i$ inside a $v$-platoon is $d_i=0$,
its new velocity is determined by $v_i'=\min(v_{i},d_{i}^{s})$ with
$d_{i}^{s}=\left[(1-\alpha )\cdot v_{p}+\frac{1}{2}\right]$.
For a stable $v$-platoon, $v_i'$ must be equal to $v$.
This gives the following stability condition:
\begin{equation}
v\leq (1-\alpha )\cdot v + \frac{1}{2}.
\label{stabcond}
\end{equation}
Equation (\ref{stabcond}) can be regarded as a condition for the
anticipation parameter $\alpha$. It implies that a $v$-platoon can 
only be stable for
\begin{equation}
\alpha_{v+1} < \alpha \leq \alpha_v
\label{vplatoonstability}
\end{equation}
where $\alpha_v$ is defined by
\begin{equation}
\alpha_v := \frac{1}{2v}.
\end{equation}
However, this condition is only necessary, not sufficient. The $v$-platoons
that can be realized for a given $\alpha$ also depend on the randomization
$R$. E.g.\ for $R=0.2$, platoons with $v=0,1,2,3$ occur, whereas for
$R=0.4$ platoons with $v=3$ can not be observed in the infinite
system although they might exist in small systems. Simulations indicate
that the slope of the mixed branch in the fundamental diagram has to
be smaller than $(1-R)(v_{f}-1)$ where $v_f$ is the average velocity in 
free flow: $v_f=v_{max}-R$. 
This will be discussed in Sec.~\ref{sub-stability} in more detail.

Another criterion for the stability of platoons can be derived
from the condition that the inflow and outflow of the platoon have
to be identical in the steady state. In the following we will derive 
estimates for these flows and in this way obtain the fundamental diagram 
in the mixed region.

The outflow from a $v$-platoon is determined by the average time $T_w$
needed by the leading vehicle of the platoon to accelerate to 
velocity $v+1$. Assuming that this car has a large headway,
this time is determined by the randomization constant $R$ through
$T_w=\frac{1}{1-R}$.
Therefore, in the free-flow region of the system, the average headway
$\Delta x_f$ is given by $\Delta x_f = T_w (v_f-v) +1$. This consideration
is very similar to the reasoning used in \cite{s2s}.

Assuming that the platoon consists of $N_v$ and the
free-flow region of $N_f$ vehicles, we have
\begin{equation}
N=N_v+N_f,\qquad {\rm and}\qquad L=N_v + N_f \Delta x_f.
\end{equation}
Here $N$ is the total number of cars. Furthermore it has been assumed
that the transition region between the platoon, where all cars have
headway $d_i=0$, and the free-flow region, where the average headway
is given by $\Delta x_f$, can be neglected. Eliminating $N_f$ we
find
\begin{equation}
\frac{N_v}{L} = \frac{\rho\Delta x_f - 1}{\Delta x_f - 1}.
\label{eq_platoonsize}
\end{equation}

We now can calculated the flow $J=\rho \bar{v}$ of the corresponding 
phase-separated state. The average velocity $\bar{v}$ in the presence
of a $v$-platoon is given by 
\begin{equation}
\bar{v} = \frac{N_v v + N_f v_f}{N}.
\end{equation}
A straightforward calculation using the results given above yields
for the flow
\begin{equation}
J_v= (1-R) + \left(v-(1-R)\right)\rho.
\label{eq_mixedflow}
\end{equation}
These results are in excellent agreement with the results from 
computer simulations. Since $1-R < 1$ it implies that all slopes
corresponding to mixed states in the fundamental diagram are positive,
except those for $0$-platoons 
($\alpha > 0.5 $ see Fig. \ref{f3b})
which are responsible for the 
jammed branch with low flow and negative slope. 


\subsection{Structure of the congested states}
\label{sub-congested}

For large densities all 
anticipatory
curves 
($ \alpha\leq 0.5$)
collapse on one congested branch where
the flow decreases with increasing density.
Simulations indicate that the structure of the corresponding states
depends on the parameter regime. In the range (\ref{vplatoonstability}), 
where a $v$-platoon can exist, the congested branch is characterized by 
the coexistence of a compact jam ($0$-platoon) and various $v$-platoons. 
The $v$-platoons are formed when a bunch of vehicles escapes from the jam. 
As argued in Sec.~\ref{sub-mixed}, the first car escapes after an average 
waiting time $T_w=\frac{1}{1-R}$. Due to anticipation, with probability 
$1-R$ the second car can move in the same time-step, and so on. 
Therefore the average number of cars escaping in the same time-step 
is given by
\begin{equation}
\bar{l}=\frac{\sum_{l=1}^\infty l(1-R)^l}{\sum_{l=1}^\infty (1-R)^l}
=\frac{1}{R}.
\end{equation}
These cars form a $v$-platoon of length $\bar{l}$ where the value
of $v$ depends on the parameter region as discussed above.
Since the average waiting time for the escape of a car is $T_w$, the
average distance between two $v$-platoons is 
$\Delta x_c = vT_w = \frac{v}{1-R}$.

To calculate the flow in the congested branch, we again neglect the
transition regions and assume that only one jam with $N_0$ vehicles
and $n$ $v$-platoons with a total number of $N_v$ cars are present.
Then we have $N_0+N_v=N$ with $N_v=n\bar{l}$. Furthermore 
$N_0+N_v+n\Delta x_c =L$ where $N_0$ and $N_v$ are the total lengths
of the platoons and $n\Delta x_c$ is the total space between the platoons.
These relations yield
\begin{equation}
1 = \frac{1}{L}\left(N+n\Delta x_c\right)=\rho + \frac{N_v}{L}\cdot
\frac{\Delta x_c}{\bar{l}}.
\label{congest1}
\end{equation}
The average velocity of the vehicles in the congested branch is 
$\bar{v}=\frac{N_v v}{N}$. Using (\ref{congest1}), this implies for 
the flow 
\begin{eqnarray}
J_{\text{cong}}&=&\rho\bar{v}=\frac{N}{L}\frac{N_v}{N}v
=(1-\rho)\frac{\bar{l}}{\Delta x_c}v  \nonumber\\
&=& \frac{1-R}{R}(1-\rho).
\label{congflow}
\end{eqnarray}
Note that this result is independent of the velocity $v$ of the
platoons! It is in excellent agreement with the simulation data,
justifying e.g.\ the assumption made about the transition regions.


\subsection{Stability regions}
\label{sub-stability}

For fixed $\alpha$ we now can estimate the stability region 
$\rho_1 \leq \rho \leq \rho_2$ for the mixed states. At the lower 
boundary density $\rho_1$ the number of cars $N_v$ in the $v$-platoon 
vanishes. From (\ref{eq_platoonsize}) one has $\rho_1\Delta x_f - 1 =0$
which yields
\begin{equation}
\rho_1 = \frac{1-R}{v_{f}-v+(1-R)}.
\label{eq_rho1}
\end{equation}
The upper bound $\rho_2$ is not determined by the condition $N_v=N$,
i.e.\ all cars belong to the $v$-platoon. This would correspond
to the density $\rho=1$. In fact, the instability of the mixed state 
occurs earlier. At the density
\begin{equation}
\rho_2  = \frac{(1-R)^2}{R(v+R-2) +1}
\label{eq_rho2}
\end{equation}
the flow (\ref{eq_mixedflow}) of the mixed branch becomes larger than 
that of the congested branch, see (\ref{congflow}) and therefore
(at least for random initial conditions) the flow of the congested
branch is observed. However, our simulations have given indications 
for hysteresis effects and metastability in the large density regime. 
We will discuss these in more detail in a future publication \cite{inprep}.
For $v=0$ the upper transition density becomes $\rho_2=1$, independent
of $R$, consistent with the observation (Fig.~\ref{f3b})
that the mixed region for $v=0$ extends up to the maximal density.

Since $\rho_2$ has to be larger than $\rho_1$, this yields an
additional condition for the stability of the branches. It is
easy to check that $\rho_1< \rho_2$ if
\begin{equation}
(1-R)v_f  > v .
\label{Rcondition}
\end{equation}
This is just the condition obtained in Sec.~\ref{sub-mixed} from
computer simulations.

Summarizing, a mixed region with $v$-platoons can only exist for
$1/(2(v+1)) < \alpha \leq 1/(2v)$ and $R$ satisfying (\ref{Rcondition}).
If these conditions are fulfilled, $v$-platoons occur in the density 
interval $\rho_1\leq \rho\leq\rho_2$ where $\rho_1$ and $\rho_2$ are 
given by (\ref{eq_rho1}) and (\ref{eq_rho2}), respectively.


\section{Summary and conclusions}
\label{summary}

Forecasting the impact of different anticipation schemes plays a
essential role in real traffic flow in order to propose automated
traffic alternatives. In this paper we have introduced and
investigated a modification of the NaSch model to better capture
reactions of the drivers intended to keep safety on the highway.  
As a result, an anticipation parameter $\alpha \in [0,1]$, that allows
to determine a velocity-dependent safe braking distance of the
precedent vehicle, is included. The addition of this parameter proves
to be useful to describe different traffic situations of
non-automated, automated, and mixed traffic.

Simulation results presented here for homogeneous drivers,
corresponding to a cautious estimation of the preceding car velocity
(large $\alpha$), are in excellent agreement with the shape of the
empirical fundamental diagram.

On the other hand, simulation results for driving schemes associated to
intermediate-levels of anticipation with $\alpha$ from $0.13$ to $0.5$
(homogeneous drivers), exhibit phase separation in a certain density
regime into a free-flow region and so-called $v$-platoons. In these
dense platoons vehicles move with the same velocity $v$ and have 
vanishing headway. The velocity $v$ of the platoon is determined
by the level of automation.  

This platoon formation observed in a mixed regime plays an
important role in Automated Highway Systems (AHS) to increase the
vehicular capacity. Therefore, results obtained help to elucidate the
effects of anticipation coded in $\alpha$. The maximum flow and the
density interval for free-flow regime go with the inverse of $\alpha$:
smaller values of $\alpha$, (greater estimation of the precedent car
velocity) imply larger flows and larger density interval for free-flow
regime. This is also in accordance with, for example, the use of
certain anticipation strategies to exploit the knowledge of the
velocity of the precedent vehicle and so, reducing the distance among
vehicles, increasing the capacity, and the density interval for
free-flow regime.

Moreover, the analysis of the speed variance of individual vehicles
also indicates the importance to anticipation effects.  Results
indicate that level of anticipation should be determined based on the
density.  The highest-level of anticipation should be considered
before the corresponding maximum density for free-flow regime is
reached.  After that maximum density, larger density requires
lower-level of anticipation.  Therefore, this selection not only
produces the highest flow, but also the lowest standard deviation, and
so, efficiency is the highest.
  
The considerations in this paper show the flexibility of the CA
approach to more complex traffic flow problems. A simple and natural
modification of the rules of the NaSch model to consider different
driving schemes allows us to describe the formation of coherently
moving platoons observed in some anticipation schemes.  We think
that the results presented here are relevant to establish suitable
levels of safety and automation not only for AHS, but also in real
traffic.  We stress that although in this paper the model is
simulated in a single-lane on a ring, it is possible to apply it to
complex highway topologies in a satisfactory way \cite{inprep}.

Apart from its practical relevance for traffic problems, our work
also shows interesting physical aspects. The model suggested here
exhibits various kinds of phase separation phenomena. At intermediate
densities, phase separation into a condensed ($v$-platoon) and a 
non-condensed (free-flow) phase can be observed. In contrast to most
other models of driven diffusion, the condensed phase is moving
coherently for $v>0$. At high densities an even more surprising
state is found that exhibits phase separation between different
condensates, a non-moving one ($v=0$) and several coherently moving
platoons ($v>0$). To our knowledge such a behavior has not been
observed before. It would be interesting to study these phases
in more detail, especially since recently some progress in the
understanding of phase separation in driven diffusive models
has been made \cite{phasesep}. Work in this direction is currently
in progress \cite{inprep}.



\begin{thebibliography}{99}

\bibitem{EKV95}  F. Eskafi, D. Khorramabadi, and P. Varaiya, 
Transpn. Res. C 3A(1), 1 (1995).

\bibitem{GETRAM}  Transport-Simulation-Systems GETRAM. Available at 
{\tt http://www.aimsun.com} (2002).

\bibitem{autobahn} Verkehrsinfo NRW.
Available at {\tt http://www.autobahn.nrw.de/}.

\bibitem{a5}  K. Nagel and M. Schreckenberg, J. Physique I 2, 2221 (1992).

\bibitem{ito} M.~Schreckenberg, A.~Schadschneider, K.~Nagel, and N.~Ito,
Phys.\ Rev.\ E 51, 2939 (1995)

\bibitem{San}  D. Chowdhury, L. Santen, and A. Schadschneider, 
Phys.\ Rep.\ 329, 199 (2000).

\bibitem{nagelreview} K. Nagel, J. Esser, and M. Rickert, 
Annu.\ Rev.\ Comp.\ Phys.\ 7, 151, ed.\ D.\ Stauffer (World Scientific, 2000)

\bibitem{es}  J. Esser and M. Schreckenberg, 
Int. J. Mod. Phys. C8, 1025 (1997).

\bibitem{and} A. Schadschneider, Physica A 285, 101 (2000);
ibid A 313, 153 (2002).

\bibitem{a2}  K. Nagel, Phys. Rev. E 53, 4655 (1996).

\bibitem {ker} B. S. Kerner, Phys. World 8, 25 (1999).

\bibitem{knospe2lane} W. Knospe, L. Santen, A. Schadschneider, and
M. Schreckenberg, Physica A265, 614 (1999).

\bibitem{knospe}  W. Knospe, L. Santen, A. Schadschneider, and
M. Schreckenberg, J. Phys. A 33, 477 (2000).

\bibitem{chen} H.-J.\ Chen, Int. J. Mod. Phys. B 15, 3453 (2001).

\bibitem{LWK}  X. Li, Q. Wu, and R. Jiang, Phys. Rev. E 64, 066128 (2001).

\bibitem{schneider} J. Schneider and  A. Ebersbach, 
Int. J. Mod. Phys. C 13, 107 (2002)

\bibitem{knospe2}  W. Knospe, L. Santen, A. Schadschneider, and
M. Schreckenberg, Phys. Rev. E 65, 015101(R) (2002).

\bibitem{wagner} N. Eissfeldt and P. Wagner, 
Eur.\ Phys.\ J.\ B 33, 121 (2003).

\bibitem{AH-JVSD99a} L. Alvarez and R. Horowitz, Veh. Sys. Dyncs. 32,
  23 (1999).

\bibitem{Larraga} M.E. L\'arraga, J.A. del R\'{\i}o, and A. Mehta, 
Physica A 307, 527 (2002).

\bibitem{SCHI94}  D. Swaroop, C. Chien, J. Hedrick, and P. Ioannou, 
Veh. Sys. Dyncs. 23 (1994).

\bibitem{SCA}  J. Rillings, Sci. Amer. 365, 60 (1997).

\bibitem{Hall}  F.L. Hall, L.A. Brian, and M.A. Gunter, 
Transpn. Res. A 20A, 197 (1986).

\bibitem{Eme}  H. Emmerich and E. Rank, Physica A 234, 676 (1997).

\bibitem{s2s} R. Barlovic, L. Santen, A. Schadschneider, and 
M. Schreckenberg, Eur.\ Phys.\ J.\  5, 793 (1998).

\bibitem{montroll} H. Reiss, A.D. Hammerich, and E.W. Montroll, 
J. Stat. Phys. 42, 647 (1986).

\bibitem{inprep} M.E. L\'{a}rraga, J.A. del R\'{\i}o, and A. Schadschneider,
in preparation.

\bibitem{phasesep} Y.\ Kafri, E.\ Levine, D.\ Mukamel, G.M.\ Sch\"utz, and
J.\ Torok, Phys. Rev. Lett. 89, 035702 (2002).
\end{thebibliography}
\end{document}